\documentclass{aa}  
\usepackage{natbib,twoopt}
\usepackage{longtable}
\usepackage{pdflscape} 
\usepackage{afterpage} 
\usepackage[breaklinks=true]{hyperref} 
\bibpunct{(}{)}{;}{a}{}{,}             
\usepackage{graphicx}
\usepackage{txfonts}
\usepackage{hyperref}
\hypersetup{
     colorlinks = true,
     linkcolor = blue,
     anchorcolor = blue,
     citecolor = blue,
     filecolor = blue,
     urlcolor = blue
     }
\begin{document}

   \title{The $\mathrm{H_2}$ angular momentum - mass relation of local disc galaxies}


   \author{Nikki N. Geesink
          \inst{1}\fnmsep\thanks{\email{geesink@strw.leidenuniv.nl}},
          Pavel E. Mancera Pi\~{n}a\inst{1}\fnmsep\thanks{\email{pavel@strw.leidenuniv.nl}}, Claudia del P. Lagos,\inst{2,3} \and Mariska Kriek\inst{1}}
    \institute{Leiden Observatory, Leiden University, P.O. Box 9513, 2300 RA, Leiden, The Netherlands
    \and 
    International Centre for Radio Astronomy Research (ICRAR), M468, University of Western Australia, 35 Stirling Hwy, Crawley, WA 6009, Australia
    \and 
    ARC Centre of Excellence for All Sky Astrophysics in 3 Dimensions (ASTRO 3D)
 }


   \date{Received ; accepted }

\abstract
{We present an analysis of the molecular specific angular momentum--mass ($j_{\rm H_2}-M_{\rm H_2}$) relation using a sample of 51 nearby disc galaxies from the PHANGS-ALMA survey with deep high-resolution molecular gas rotation curves and surface density profiles. For the very first time, using a statistical sample, we report the discovery of a well-defined $j_{\rm H_2}-M_{\rm H_2}$ relation. We quantify the scaling law by fitting a  power law with a Bayesian framework, finding $j_{\rm H_2}\propto M_{\rm H_2}^{0.53}$. This slope closely resembles the well-known stellar $j_{\ast}$–$M_{\ast}$ (Fall) relation, highlighting the dynamical connection between molecular gas and stars. We show that the $j_{\rm H_2}-M_{\rm H_2}$ relation cannot be fully explained by analytic models of disc stability, but instead is well recovered with more complex physics as implemented in the \textsc{Shark} semi-analytical model. These findings demonstrate the power of our novel $j_{\rm H_2}-M_{\rm H_2}$ relation in testing galaxy evolution theories and in setting new constraints for models and simulations that  reproduce a realistic interstellar medium. Additionally, our findings provide a critical benchmark for upcoming high-redshift studies of molecular gas kinematics, offering a local baseline to study the evolution of cold gas dynamics across cosmic time.   }


\titlerunning{The $\mathrm{H_2}$ angular momentum–mass relation of local disc galaxies}
\authorrunning{N.N. Geesink et al.}
   \maketitle
%
\section{Introduction}
Specific angular momentum ($j=J/M$) and mass ($M$) are two of the most fundamental properties describing any physical system, including galaxies (e.g. \citealt{white_angular_1984,fall_formation_1980, romanowsky_angular_2012, cimatti_introduction_2019}). Angular momentum in galaxies is thought to be acquired by gravitational tidal torques before virialisation, according to the tidal torque theory \citep{peebles_origin_1969, white_angular_1984}. This theory predicts a scaling relation for the specific angular momentum of the dark matter haloes of the form  $j_{\rm DM} \propto M_{\rm DM}^{2/3}$ \citep{fall_galaxy_1983, shaya_angular_1984, heavens_tidal_1988}. Baryonic matter is expected to be subjected to the same tidal torques as the dark matter haloes since during the linear stage of structure formation the dark matter and primordial gas are still well mixed. 

Observationally,
\cite{fall_galaxy_1983} first showed that both early- and late-type galaxies follow a scaling law in the stellar angular momentum ($j_\ast$) versus stellar mass ($M_\ast$) plane,  called the Fall relation. Even though the normalisation  for spiral galaxies is higher than that for early types, both classes follow a relationship of the form $j_{\rm \ast} \propto M^{\alpha}_{\ast}$, with $\mathrm{\alpha \approx 0.6}$. Later studies with more and better data confirmed the results of \cite{fall_galaxy_1983}, generally finding a power law with a slope of around 0.5-0.6 \citep{romanowsky_angular_2012, fall_angular_2018, posti_angular_2018, mancera_pina_baryonic_2021}, consistent (within the uncertainties) with the expectations from tidal torque theory for the dark matter haloes.

In addition to the stellar component, the angular momentum of the cold gas is also vital for galaxy evolution, but has been studied significantly less. In recent years, the neutral atomic hydrogen (H\,{\sc i}) $j_{\rm HI}-M_{\rm HI}$ relation has started to be studied using resolved interferometric observations of late-type massive and dwarf galaxies \citep{cortese_sami_2016, chowdhury_angular_2017, kurapati_angular_2018, mancera_pina_baryonic_2021, mancera_pina_tight_2021}. This $j_{\rm HI}-M_{\rm HI}$ relation is found to be significantly steeper than the Fall relation, $j_{\rm HI}\propto M_{\rm HI}$ \citep{kurapati_angular_2018, kurapati_h_2021,mancera_pina_baryonic_2021}. 
In contrast, the picture for the molecular gas (H$_2$) is largely incomplete due to the scarcity of deep and high-resolution CO data. In fact, H$_2$ is often neglected when building the baryonic $j-M$ relation, $j_{\rm bar}\propto M_{\rm bar}^{0.6}$  (e.g. \citealt{elson_relation_2017, kurapati_angular_2018, murugeshan_influence_2020, mancera_pina_baryonic_2021}). \cite{obreschkow_fundamental_2014} estimated $j_{\rm H_2}$ for a small sample of 16 nearby galaxies from the THINGS survey \citep{walter_things_2008}. However, they did not quantify the shape of the $j_{\rm H_2}-M_{\rm H_2}$ relation, and their measurements show large scatter, which casts doubts on the existence of a relation.

Considering that angular momentum regulates galaxy sizes, morphologies, and gas content \citep{fall_galaxy_1983, mo_formation_1998, romanowsky_angular_2012, pezzulli_accretion_2016,mancera_pina_tight_2021, hardwick_xgass_2022, elson_existence_2024} and that molecular gas is the primary fuel for star formation \citep{kennicutt_star_1998, bigiel_star_2008}, lacking a quantification of the $j_{\rm H_2}-M_{\rm H_2}$ relation is a significant gap in our attempts to understand the physical properties of the interstellar medium (ISM) in galaxies through their evolutionary pathways \citep{lilly_gas_2013, lagos_angular_2017, catinella_xgass_2018, tacconi_evolution_2020, saintonge_cold_2022}. Moreover, since processes such as gas accretion, feedback, mergers, and dynamical friction can dramatically alter a galaxy’s angular momentum distribution \citep{teklu_connecting_2015, stevens_connecting_2018, sweet_stellar_2020}, quantifying the molecular $j-M$ relation can also offer important constraints for testing and refining models and simulations of galaxy evolution. 

In addition, characterising the $j_{\rm H_2}-M_{\rm H_2}$ relation at $z=0$ is a crucial benchmark for high-redshift observations. To date, only $j_\ast$ has been studied up to $z \approx 1-2$ (see \citealt{marasco_angular_2019, sweet_angular_2019, bouche_muse_2021, espejo_salcedo_multiresolution_2022, mercier_stellar_2023}); there is no consensus on its evolution. 
The molecular phase is vital in order to study the gas counterpart since H\,{\sc i} can only be detected in emission up to $z \lesssim 0.2$ \citep{gogate_budhies_2020, ponomareva_mightee-h_2021}. For example, thanks to ALMA, it is now becoming common to trace the molecular gas kinematics and distribution at $z\gtrsim3$ (e.g. \citealt{rizzo_dynamically_2020, rizzo_alma-alpaka_2023, rowland_rebels-25_2024}). To benchmark the angular momentum content in those young galaxies and their evolution through time, determining the local $j_{\rm H_2}-M_{\rm H_2}$ is imperative.  
In this paper, we exploit recent deep high-resolution molecular gas observations of 51 local disc galaxies from the PHANGS--ALMA survey
to characterise the $j_{\rm H_2}-M_{\rm H_2}$ relation for the first time. 

This paper is organised as follows. Section~\ref{sec:galsample} describes our galaxy sample, and Sect.~\ref{sec: fitting} outlines the methods for determining the specific angular momentum. Our results are presented in Sect.~\ref{sec:jmrela} and Sect.~\ref{sectestmodel}. Finally, we summarise our conclusions in Sect.~\ref{secconclusion}.
Throughout the text, we adopt a $\Lambda$CDM cosmology with  H$_0 = 70~{\rm km~s}^{-1}~{\rm Mpc}^{-1}$, $\Omega_{\rm m} = 0.3$, and  $\Omega_{\Lambda} = 0.7$.

\section{Galaxy sample}
\label{sec:galsample}
Our sample was drawn from the PHANGS-ALMA survey \citep{leroy_dataphangs-alma_2021, leroy_surveyphangs-alma_2021}, mapping CO $J = 2 \rightarrow 1$ emission, hereafter CO($2-1$), in 90 nearby ($d \lesssim 20\, \rm Mpc$) star-forming galaxies at $\sim 1^{\prime\prime}$ resolution. As detailed below, in selecting our sample, we included galaxies with available CO rotation curves and surface density profiles derived by the PHANGS-ALMA collaboration. 

\subsection{Rotation curves}
\label{sec:rotationcurves}
 We used the rotation curves derived by \cite{lang_phangs_2020} for the PHANGS-ALMA sample, obtained using the  commonly employed tilted-ring model of the observed velocity field \citep{rogstad_aperture-synthesis_1974, bosma_distribution_1978, begeman_hi_1987}.  \cite{lang_phangs_2020} provided high-resolution (150 pc) rotation curves for 67 galaxies with ordered rotation. 

To improve the radial coverage of the kinematics of our galaxy sample, we supplemented the CO rotation curves with literature H\,{\sc i} rotation curves when available. 
H\,{\sc i} extends  well beyond the optical and CO emission of galaxies, making it an effective tracer for galaxy dynamics at the outer radii;  we note that H\,{\sc i} and CO are expected to corotate, and this has  been corroborated with observations in nearby galaxies \citep{bacchini_volumetric_2020, laudage_neutral_2024}. Specifically, we supplemented the CO rotation curves of NGC~2903, NGC~3351, NGC~3521, and NGC~3627, NGC~1365, NGC~3621, NGC~4535, and NGC~4536,  with H\,{\sc i} rotation velocities (also from tilted-ring models) from \cite{di_teodoro_radial_2021} and \cite{ponomareva_detailed_2016}. In addition, we included H\,{\sc i} rotation curves for three galaxies without available CO rotation curves, bringing the total sample to 70 galaxies. We obtained the H\,{\sc i} rotation curves for NGC~253 from \cite{mancera_pina_impact_2022}, NGC~300 from \cite{mancera_pina_baryonic_2021}, and NGC~7793 from \cite{bacchini_volumetric_2019}. Where needed, we   corrected the rotation curve data to match the distance and inclination  reported in \cite{lang_phangs_2020} to ensure homogeneity. 

\subsection{Gas surface densities}
We used the molecular gas surface density profiles provided by \cite{sun_molecular_2022} for 66  of our 70 galaxies. The molecular gas surface density profiles ($\Sigma_{\rm H_2}$) were derived from the integrated CO(2-1) line intensity $I_{\rm CO}^{(2-1)}$ via
\begin{align}
    \frac{\Sigma_{\text{H}_2}}{M_{\odot} \, \text{pc}^{-2}} &= \frac{\alpha_{\text{CO}}^{(1-0)}}{M_{\odot} \, \text{pc}^{-2}} \frac{1}{R_{21}} \frac{I_{\text{CO}}^{(2-1)}}{\text{K km/s}},  
\label{eq:surfdens}
\end{align}
where $R_{21} = 0.65$ is the adopted CO(2-1) to CO(1-0) line ratio \citep{den_brok_new_2021, leroy_low-j_2022}, and ${\alpha_{\text{CO}}^{(1-0)}}$ is the CO-to-$\rm H_2$ conversion factor, for the CO(1-0) line.

The quantity $\alpha_{\rm CO}^{(1-0)}$ remains challenging to constrain precisely due to its sensitivity to local conditions such as gas density, temperature, and metallicity, which vary significantly across galactic environments \citep{schinnerer_molecular_2024}. The database of \cite{sun_molecular_2022} provides four alternative prescriptions for the PHANGS-ALMA sample, considering various factors that impact ${\alpha_{\text{CO}}^{(1-0)}}$. 
\begin{enumerate}
    \item The fiducial conversion factor used by \cite{sun_molecular_2022}, which follows a metallicity-dependent calibration first introduced by \citet[][see also e.g. \citealt{accurso_deriving_2017, schinnerer_molecular_2024}]{sun_dynamical_2020}:
\begin{align}
\alpha_{\text{CO, S20}} &= 4.35 \, Z'^{-1.6} \, M_{\odot} \, (\text{K km s}^{-1} \, \text{pc}^2)^{-1}.
\label{eq:alpha_co_sun}    
\end{align}

\item The commonly used Galactic average value (see \citealt{bolatto_co--h2_2013, sandstrom_co--h_2013}): 
\begin{align}
    \alpha_{\text{CO, MW}}= 4.35 \, M_{\odot}\, \mathrm{(K\,km\,s^{-1}\, pc^{2})}^{-1}~.
\end{align}

\item A calibration that takes into account both metallicity and line intensity from the numerical work by \cite{narayanan_general_2012}:
\begin{align}
    \frac{\alpha_{\text{CO, N12}}}{M_{\odot} \, \text{pc}^{-2} \, (\text{K km s}^{-1})^{-1}} &= 8.5 \, Z'^{-0.65} \, \min\left[1, \, 1.5 \times \left( \frac{\left\langle I_{\text{CO}(2-1)} \right\rangle}{\text{K km s}^{-1}} \right)^{-0.32} \right]~.
\label{eq:alpha_co_N12}
\end{align}
Here the dependence on the line intensity follows from the effect of varying gas temperature and velocity dispersion in galaxies. 

\item Following \cite{bolatto_co--h2_2013}, a calibration considering a dependence on the molecular cloud surface density and total baryonic surface density:
\begin{align}
    \frac{\alpha_{\text{CO, B13}}}{M_{\odot} \, \text{pc}^{-2} \, (\text{K km s}^{-1})^{-1}} &= 2.9 \exp \left( \frac{40 \, M_{\odot} \, \text{pc}^{-2}}{Z' \left\langle \Sigma_{\text{mol, pix}} \right\rangle} \right) \left( \frac{\Sigma_{\text{total}}}{100 \, M_{\odot} \, \text{pc}^{-2}} \right)^{-\gamma} \\
    \text{with} \quad \gamma &= 
    \begin{cases} 
      0.5, & \text{if } \Sigma_{\text{total}} > 100 \, M_{\odot} \, \text{pc}^{-2} \\
      0, & \text{otherwise}
    \end{cases}.
\end{align}
Here $\Sigma_{\text{mol, pix}}$ gives the molecular cloud surface density and $\Sigma_{\text{total}}$ the total surface density (see \citealt{sun_molecular_2022}).

\end{enumerate}

\begin{figure*}
   \centering
   \includegraphics[width=\textwidth]{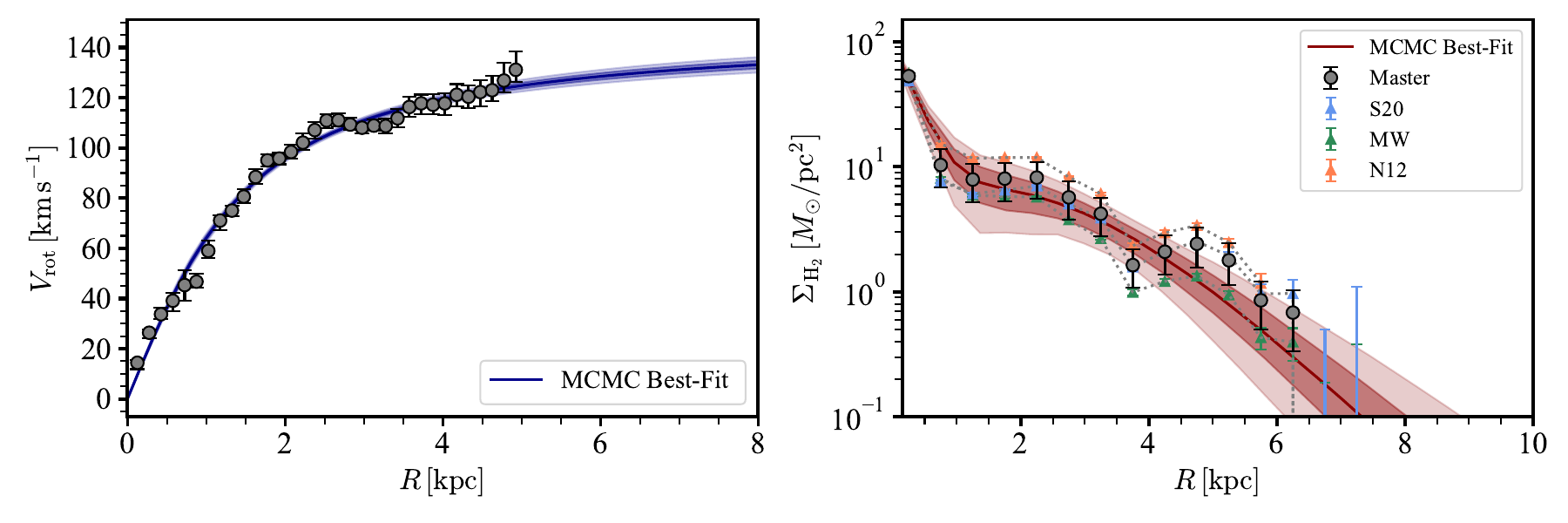}
   \caption{Example of the functional forms fitted to the rotational velocities and molecular gas surface density profiles (in this case for the galaxy IC~5273). \textit{Left:} Observed rotation curve (grey markers with error bars) and the best-fit model  (solid blue curve) with its 1$\sigma$ and 2$\sigma$ uncertainties represented by the shaded region. \textit{Right:} Molecular gas surface density profile assuming our master $\alpha_{\text{CO}}$ conversion factor (grey markers with error bars). The best-fit model is shown as a solid red curve, with its 1$\sigma$ and 2$\sigma$ uncertainties indicated by the shaded region. Profiles derived using alternative CO-to-H$_2$ conversion factors (S20, MW, N12) are overplotted for comparison (B13 was not available for IC~5273). As can be seen, our profile represents a good compromise between the different calibrations and incorporates realistic uncertainties.}
   \label{Figprofilefits}
\end{figure*}

\noindent The difference in $\alpha_{\text{CO}}$ between the different calibrations often differs by more than the typical uncertainties quoted in the above references.
To incorporate these different prescriptions to obtain $\alpha_{\text{CO}}$, we adopted the following approach. For each galaxy, at each radius, we combine the available $\alpha_{\text{CO, MW}}$, $\alpha_{\text{CO, S20}}$, $\alpha_{\text{CO, B13}}$, and $\alpha_{\text{CO, N12}}$ to construct a master conversion factor. The master conversion factor is derived by averaging the four prescriptions, and we adopt the standard deviation as uncertainty. The master $\alpha_{\text{CO}}$ is then used in Eq. \ref{eq:surfdens} to derive our final $\Sigma_{\rm H_2}(R)$ profile, with uncertainties determined through standard error propagation. Combining the calibrations into a master $\alpha_{\text{CO}}$ is a conservative approach that manages to capture realistic uncertainties better; this is demonstrated in the right panel of Fig.~\ref{Figprofilefits}, where the resulting surface density profiles for the various $\alpha_{\rm CO}^{(1-0)}$ calibrations are shown for IC~5273. \\

\noindent
From our 66 galaxies we also excluded NGC~1512, NGC~2566, NGC~2775, NGC~4569, and NGC~4826 since their surface density profiles are too irregular to allow   a robust quantification of their angular momentum. Similarly, the rotation curve of NGC~5068 is too compact and shows no signs of flattening, making it unsuitable for our purposes. 
Considering this, we ended up with a final sample of 60 galaxies, spanning the mass ranges $10^9 \lesssim M_{\ast}/M_{\odot} \lesssim 10^{11}$, $10^8 \lesssim M_{\rm H_2}/M_{\odot} \lesssim 10^{10}$, and $10^{-2} \lesssim M_{\rm H_2}/M_{\ast} \lesssim 10^{-0.75}$. In Table~\ref{tablejvals} we present the final sample with their Hubble types, stellar masses, and distances adopted from \cite{leroy_surveyphangs-alma_2021}.


\section{Computing j} 
\label{sec: fitting}

In this section we describe how we computed  the specific angular momentum for our galaxy sample by exploiting their rotation curves and surface density profiles. 
For a disc, the specific angular momentum within a radius $R$ is described by
 \begin{equation}
     j(< R) = \frac{J(<R)}{M(<R)} = \frac{2\pi \int_{0}^{R} R'^2 \Sigma(R') V(R') \, dR'}{2\pi \int_{0}^{R} R' \Sigma(R') \, dR'}\,
    \label{eq:specific j},
 \end{equation}
where $V(R)$ is the rotation curve and $\Sigma(R)$ the surface density profile. For a typical disc galaxy, most of the angular momentum can be measured by tracing the kinematics and mass surface density out to $R\geq 2 R_{\rm e}$ \citep{romanowsky_angular_2012,posti_angular_2018, mancera_pina_baryonic_2021}, where $\Sigma(R)$ becomes small and $V(R)$ has flattened. Measuring $j$ from insufficiently extended data can lead to its underestimation (see Sect. \ref{secconv}). Exploiting the empirical facts that rotation curves are flat at large radii \citep{bosma_distribution_1978, begeman_hi_1987, de_blok_high-resolution_2008, kuzio_de_naray_mass_2008} and gas surface densities decay exponentially \citep{bigiel_star_2008, wang_new_2016}, we extrapolate the observed data to capture all $j$. For this, we fit $\Sigma(R)$ and $V(R)$ with functional forms.

\subsection{Rotation curve fitting}
We considered two models to extrapolate $V(R)$: an arctan function \citep{courteau_optical_1997} and a more flexible multi-parameter function \citep{rix_internal_1997}. The functions are given, respectively, by
\begin{align}
    V_{\text{rot}}(R) &= V_0 \frac{2}{\pi} \arctan(R / r_{\rm t}),
    \label{eq: arctan}
\end{align}
\begin{align}
V_{\mathrm{rot}}(R) = V_{t} \frac{\left( 1 + \frac{r_{\rm t}}{R} \right)^{\beta}}{\left[ 1 + \left( \frac{r_{\rm t}}{R} \right)^{\xi} \right]^{1/\xi}}.
\label{eq:multi}
\end{align}
Here $V_0$ is the asymptotic velocity reached at an infinite radius, and $r_{\rm t}$ is the turnover radius between the rising and the outer flat part of the rotation curve. The $V_{\rm t}$ parameter Eq. \ref{eq:multi} is a scale velocity regulating the amplitude of the rotation curve. The two additional parameters, $\xi$ and $\beta$, regulate the rise and flattening of the rotation curve. 

For simplicity, we first fitted the rotation curves with the arctan function, finding satisfactory fits for most cases. For five galaxies (NGC~300, NGC~1365, NGC~2903, NGC~3521, and NGC~3627) with complex kinematics the multi-parameter function works better. In practice, we retrieved posterior distributions for the best-fitting parameters using a Markov chain Monte Carlo (MCMC) routine with the Python package \texttt{emcee} \citep{foreman-mackey_emcee_2013}, assuming flat priors on the fitting parameters. Our models successfully reproduced the observed kinematics, as illustrated in Fig.~\ref{Figprofilefits} for one representative galaxy. 

 

\subsection{Surface density profile fitting}
To fit $\Sigma_{\text{H}_2}(R)$, we considered a polyexponential function, which has been shown to effectively capture the complex behaviour of the gas surface densities in nearby galaxies \citep{bacchini_volumetric_2019, mancera_pina_impact_2022}. The polyexponential profile is given by
\begin{align}
    \Sigma_{\text{H}_2}(R) &= \Sigma_{0} e^{-R/R_{\Sigma}} \left(1 + c_1 R + c_2 R^2 + c_3R^3\right),
\end{align}
where $\Sigma_{0}$ is the central surface density; $R_{\Sigma}$ is the scale radius; and $c_1$, $c_2$, and $c_3$ are polynomial coefficients. For some profiles (especially those with sharp peaks or strong declines), one polyexponential profile is not flexible enough to reproduce all the observed surface density features. In those cases, we fit instead a combination (sum) of two polyexponential profiles. Similar to our approach with the rotation curves, we used \texttt{emcee} to determine the best-fitting parameters. An example of the fit to a representative $\Sigma_{\text{H}_2}(R)$ profile is shown in Fig.~\ref{Figprofilefits}.

\subsection{Specific angular momentum}
\label{secconv}
We adopted a Monte Carlo approach to compute the specific angular momentum and its uncertainties. Specifically, we generated a distribution of $j_{\rm H_2}$ values by calculating $j_{\rm H_2}$ for 100000 random MCMC realisations of both the rotation curves and surface density profiles fitting parameters. For each realisation, we integrated the angular momentum profiles (i.e. Eq.~\ref{eq:specific j}) up to 50 kpc. 

Our integration limit exploits the extrapolated profiles described in the previous section and is a choice to ensure that we obtained a converged estimation of $j_{\rm H_2}$. We note that the exact value of the upper limit has no impact on our results since $\Sigma_{\rm H_2}(R) \approx 0$ well before our integration limit. Nevertheless, galaxies for which $j_{\rm H_2}(<R)$ is not sufficiently converged at the limits of the data, typically galaxies without flattening of the rotation curve,  will have a larger dependence on the functional forms used in the extrapolation. We define a convergence factor $\mathcal{R}$ to quantify the degree of dependence of our results on our extrapolation, which allows us to avoid over-reliance on the extrapolated data. $\mathcal{R}$ is defined as the ratio of $j_{\rm H_2}(<R)$ evaluated at the extent of the rotation curve data to the extrapolated $j_{\rm H_2}(<R)$ (see e.g. \citealt{mancera_pina_baryonic_2021}). We establish a threshold such that galaxies with $\mathcal{R}<0.7$ are excluded from the derivation of the molecular $j - M$ relation (see below). In Appendix~\ref{sec: appconvergence}, we examine the impact of the minimum required convergence factor on our results, but we emphasise already that our results below are robust against sensible variations of $\mathcal{R_{\rm min}}$. 
Applying the convergence cutoff results in a final sample of 51 galaxies with converged $j_{\rm H_2}$ profiles.
Table~\ref{tablejvals} (\ref{tablenonconvjvals}) lists the $j_{\rm H_2}$ and $M_{\rm H_2}$ for our converged (non-converged) galaxies.\footnote{For consistency, $M_{\rm H_2}$ is estimated by integrating the denominator in Eq.~\ref{eq:specific j}. We find great agreement when comparing our values to the CO luminosities reported in \citep{leroy_surveyphangs-alma_2021}, with a median difference of 0.03 dex.} These values and their uncertainties (which also account for distance uncertainties as reported in Table~\ref{tablejvals}) correspond to the medians and $1\sigma$ uncertainties obtained with our Monte Carlo realisations, as described above. 

\section{The molecular $j-M$ relation}
\label{sec:jmrela}
\subsection{Shape and dependences}
\begin{figure}
   \centering
   \includegraphics[width=\hsize]{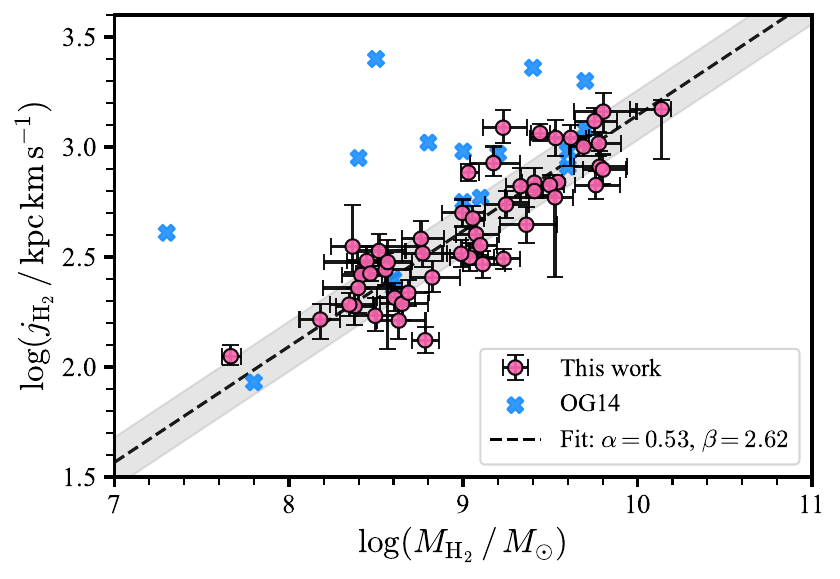}
      \caption{Molecular $j-M$ relation for our final sample of converged galaxies. The circles show our sample, while the crosses show the results of \cite{obreschkow_fundamental_2014}. The dashed line indicates the best-fitting relation, and the grey band shows its orthogonal intrinsic scatter.
              }
         \label{Fig_rela}
   \end{figure}

In Fig.~\ref{Fig_rela} we show the distribution of our converged sample in the $j_{\rm H_2}-M_{\rm H_2}$ plane. Our analysis reveals a clear relation, with a  particularly tight scaling law for $M_{\rm H_2} \gtrsim 10^9\, M_{\odot}$, and with increasing scatter below this mass. We fit the observed distribution with a power law of the form
\begin{align}
    \log\left( \frac{j_{\rm H_2}}{\text{kpc km s}^{-1}} \right) &= \alpha [\log(M_{\rm H_2} / M_{\odot}) - 9] + \beta~, 
    \label{eq: powerlawfit}
\end{align}
where $\alpha$ denotes the slope of the relation and $\beta$ the intercept. We include the orthogonal intrinsic scatter ($\sigma_{\perp}$) following \cite{bacchini_volumetric_2019} and \cite{mancera_pina_baryonic_2021}. 
We find the best-fitting parameters to be $\alpha = 0.53 \pm 0.04$, $\beta = 2.62 \pm 0.02$, with an orthogonal intrinsic scatter $\sigma_{\perp}=0.11 \pm 0.02$. In Fig. \ref{Fig_rela} we also show the best-fitting relation and its intrinsic scatter.\footnote{The inclusion of H\,{\sc i} rotation curves does not bias our results for the best-fitting relation; we verify this  explicitly in Appendix~\ref{apprc}.} The slope of the molecular relation aligns closely with that of the $j_\ast$-$M_\ast$ relation ($\alpha\approx0.5-0.6$, e.g. \citealt{romanowsky_angular_2012, fall_angular_2018, posti_angular_2018, mancera_pina_baryonic_2021}, whereas it is less steep than the $j_{\rm HI}-M_{\rm HI}$ relation ($\alpha\approx0.8-1$, \citealt{kurapati_h_2021,mancera_pina_baryonic_2021}). \\

 \noindent
With our best-fitting relation, we examined the $j_{\rm H_2}$ residuals ($\Delta_{j_{\rm H_2}-M_{\rm H_2}} = j_{\rm H_2}-j_{\rm H_{2},fit}$) as a function of various parameters to identify any underlying secondary dependences. The parameters considered include Hubble type, effective radius ($R_{\rm e}$), star formation rate (SFR), stellar mass ($M_{\ast}$), and molecular gas fraction $M_{\rm H_2} / M_\ast$. We obtained the values for Hubble type, $R_{\rm e}$, SFR, and $M_{\ast}$ from \cite{leroy_surveyphangs-alma_2021}. First, we note that we find no significant correlation between $j_{\rm H_2}$ and star formation rate (SFR) at fixed $M_{\rm H_2}$.
We surmise that this could be due to the narrow range of SFRs in our sample (by selection, the PHANGS-ALMA galaxies lie on the star-forming main sequence) or that the variations in the star formation efficiency of H$_2$ are independent of $j_{\rm H_2}$.


Among all the other parameters, the only significant trend ($p-$value=0.003) observed in our residual analysis is a moderate anti-correlation between $\Delta_{j_{\rm H_2}-M_{\rm H_2}}$ and $M_{\rm H_2}/M_\ast$. This relationship indicates that galaxies with higher $M_{\rm H_2}/M_\ast$ tend to have lower $j_{\rm H_2}$ at a given $M_{\rm H_2}$. Similarly, higher $M_{\rm H_I}/M_\ast$ are associated with lower H\,{\sc i}  specific angular momentum, while the opposite happens in the $j_\ast-M_\ast$ relation, with galaxies with high $j_\ast$ having a high $M_{\rm H_I}/M_\ast$ at fixed $M_\ast$ (see \citealt{mancera_pina_tight_2021}). We find no clear dependence with Hubble type or  $R_{\rm e}$ in our $j_{\rm H_2}-M_{\rm H_2}$ relation. While the main focus of this work is on the $H_2$ relation, we also performed preliminary explorations of the $j_{\rm H_2}-M_{\star}$ and its dependences, which we discuss in Appendix~\ref{appstellar}.

\subsection{Comparison with previous works}
In Fig.~\ref{Fig_rela} we compare our results to the study of \cite{obreschkow_fundamental_2014}, who analysed 16 nearby spiral galaxies from the THINGS survey \citep{walter_things_2008}. In their case the molecular gas surface densities were obtained from CO(2--1) maps at 11\arcsec\ resolution from the HERACLES survey \citep{leroy_heracles_2009}, when available, or from CO(1--0) maps at 7\arcsec\ resolution from the BIMA survey \citep{helfer_bima_2003} otherwise. The data used by \cite{obreschkow_fundamental_2014} was of lower resolution than those exploited in this work; in addition they used a higher line ratio of $I_{\rm CO(2 \to 1)} = 0.8 \, I_{\rm CO(1 \to 0)}$ and they adopted the $\alpha_{\text{CO}}$ of the Milky Way. For the kinematics, the approaches were also somewhat different since \cite{obreschkow_fundamental_2014} derived H\,{\sc i} rotation velocities based on a pixel-by-pixel fitting technique (see their Appendix B for details) and assumed co-rotation between H\,{\sc i} and $\mathrm{H_2}$.
As shown in Fig.~\ref{Fig_rela}, unlike our results, the sample from \cite{obreschkow_fundamental_2014} exhibits greater scatter, biased towards higher $J_{\rm H_2}$ values, and does not show a clear trend. We find that the overlapping galaxies in our samples   differ in the $j_{\rm H_2}-M_{\rm H_2}$ plane; these differences are likely attributable to the different $\alpha_{\text{CO}}$ and line ratios adopted. For these galaxies (NGC~628, NGC~3351, NGC~3521, NGC~3627, and NGC~7793), the mean (median) difference in $j_{\rm H_2}$ is 0.04 (0.13) dex, with a maximum discrepancy of approximately 0.25 dex. Whereas the difference in $j_{\rm H_2}$ is not systematic towards one direction,   for all of the overlapping galaxies (except NGC~7793) \cite{obreschkow_fundamental_2014} finds lower $M_{\rm H_2}$ values compared to our values. The mean (median) difference, without taking NGC~7793 into account, is 0.32 (0.28) dex, with a maximum difference of about 0.7 dex. For NGC~ 7793, \cite{obreschkow_fundamental_2014} find a higher value of $M_{\rm H_2}$; however, this is the only galaxy in their sample for which they infer $M_{\rm H_2}$ from the SFR (see their Sect. 2.2) instead of from the CO maps of \cite{leroy_star_2008}. \\

\noindent
Overall, our results show the close similarity between the $j_{\rm H_2}-M_{\rm H_2}$ and $j_{\ast}-M_{\ast}$ relations and highlight the significance of molecular angular momentum as an important regulator of the interstellar medium. Future work with samples that span a larger range of physical properties, such as the upcoming KILOGAS ALMA survey, will shed light on possible dependences and further refine our understanding of the $j_{\rm H_2}$–$M_{\rm H_2}$ relation. Despite this, our current relation can already be used to test theoretical models, and in the next section we provide some first examples.

\section{Testing theoretical models}
\label{sectestmodel}
Angular momentum measurements (to date only $j_\ast$ and $j_{\rm HI}$) have been used to test and constrain analytic models based on disc stability (e.g. \citealt{obreschkow_angular_2016,romeo_massive_2020, romeo_specific_2023}) as well as more elaborated semi-analytical models and hydrodynamical simulations (e.g. \citealt{obreja_nihao_2016, stevens_building_2016,lagos_angular_2017,el-badry_gas_2018, zoldan_structural_2018, lagos_quenching_2024}). Our novel work allows us to exploit the molecular $j$-$M$ relation for the first time. In this section we compare our results against the expectations of an analytic stability model and a complex semi-analytic model.

First, we compare our findings to the scaling relation driven by disc instability introduced by \cite{romeo_massive_2020} as $\frac{j_i \hat{\sigma}i}{G M_i} \approx 1$, where $\hat{\sigma}$ is the radial velocity dispersion, properly averaged and rescaled, and $i$ denotes a given mass component. This relation appears to be in good agreement with observations of the H\,{\sc i} phase (\citealt{mancera_pina_baryonic_2021}, but see also \citealt{mancera_pina_tight_2021}). For H$_2$, we can rewrite the expression by \cite{romeo_massive_2020} as
\begin{align}
    j_{\rm H_2} \propto \frac{M_{\rm H_2}}{\hat{\sigma}_{\rm H_2}}~,
\end{align}
where $\hat{\sigma}_{\rm H_2} = 0.4\bar{\sigma}_{\rm H_2}$, and $\bar{\sigma}_{\rm H_2}$ is the radial average of the $\mathrm{H_2}$ velocity dispersion (see Eq. 4 and Sect. 2.3 of \cite{romeo_massive_2020}). To match this stability relation to our observed $j_{\rm H_2}$–$M_{\rm H_2}$ relation with a slope $\alpha\approx0.5$, we need the scaling $\sigma \propto M_{\rm H_2}^{0.5}$ (but see also \cite{romeo_lenticulars_2020} for potential second-order dependences ). Instead, for our data, the dependence is much weaker, $\sigma \propto M_{\rm H_2}^{\gamma}$, with $\gamma\approx0.05-0.2$, as shown by the kinematic measurements by \cite{sun_molecular_2022, rizzo_alma-alpaka_2024}. 

We infer that models purely based on disc stability, although powerful tools with the virtue of being simple enough to be falsifiable and intuitive, appear to only partially capture the $j_{\rm H_2}-M_{\rm H_2}$ relation. 
We surmise that there could be different reasons for this shortcoming. On the one hand, numerous other factors beyond stability conditions can influence the distribution of angular momentum in galaxies, such as gas accretion, star formation, feedback, mergers, and dynamical friction (e.g. \citealt{pezzulli_angular_2017, lagos_angular_2017,stevens_connecting_2018,cimatti_introduction_2019}). In addition, it could be that the limitations in the model arise from its attempt to globalise the Toomre parameter (a local condition at a given radius) into one single average value (see \citealt{romeo_massive_2020} for details). Finally, we also note that the \cite{romeo_massive_2020} relation was calibrated with the data from \cite{obreschkow_fundamental_2014}, which shows a scattered $j_{\rm H_2}-M_{\rm H_2}$ plane that does not fully conform with our new results.

For the semi-analytic models, we performed a first exploration using \textsc{Shark} \citep{lagos_quenching_2024}. \textsc{Shark} (v2.0) incorporates a range of physical processes, including halo growth and mergers, gas accretion, chemical enrichment, and stellar and AGN feedback. In addition, \textsc{Shark} features a direct and independent modelling of $j_\ast$, $j_{\rm HI}$, and $j_{\rm H_2}$. \textsc{Shark} has demonstrated great consistency with the observed scaling relations, such as those between SFR, $M_\ast$, gas content, specific SFR, and black hole mass, making it a valuable model to compare our findings to.

  \begin{figure}[!t]
   \centering
   \includegraphics[width=\hsize]{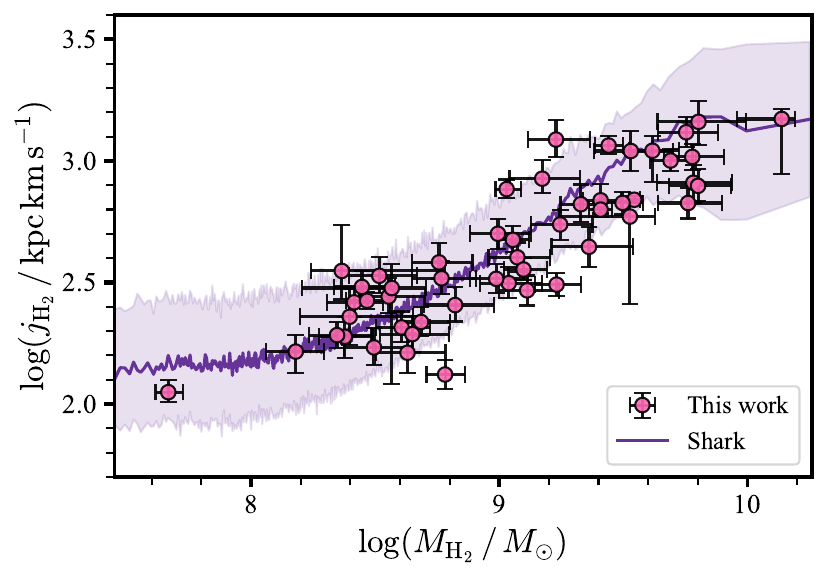}
      \caption{Comparison between our measured $j_{\rm H_2}-M_{\rm H_2}$ relation and the prediction of the \textsc{Shark} semi-analytical model (version 2.0). The \textsc{Shark} sample was selected to include galaxies with $M_\ast$ and $B/T$ in the range of the observational data. The median distribution of \textsc{Shark} is shown with a solid curve, and the 1$\sigma$ uncertainty is given by the purple band.}
         \label{Fig_shark}
   \end{figure}

To compare our $j_{\rm H_2}-M_{\rm H_2}$ relation with \textsc{Shark}, we selected only \textsc{Shark} galaxies with $M_\ast$ within $\pm 3 \sigma$ of our own $M_\ast$ range and $B/T<0.3$, ensuring a focus on disc galaxies in a comparable mass regime. Figure~\ref{Fig_shark} presents our comparison, which shows a remarkable agreement between the \textsc{Shark} predictions and our data within our observed masses and $B/T$ ranges. 
Overall, the agreement in Fig.~\ref{Fig_shark} suggests that the physical processes incorporated in \textsc{Shark} (see \citealt{lagos_shark_2018,lagos_quenching_2024} for details) manage to capture the underlying mechanisms shaping the $j_{\rm H_2}-M_{\rm H_2}$ relation at $z=0$. 

We note that at the low- and high-mass regimes \textsc{Shark} predicts a flattening of the relation.\footnote{Some models and simulations suggest a similar flattening or break at the low-mass end of the $j_{\star}-M_{\star}$ relation \citep{obreja_nihao_2016, stevens_building_2016}, although it has not been seen in observations \citep{mancera_pina_baryonic_2021}} It will be interesting to test this prediction with larger galaxy samples, even though this will be complicated for the low-mass end, given the low metallicity of low-mass galaxies. Additional future promising research avenues include a dedicated study of different hydrodynamical simulations (which will allow us to test whether the physics implementation of \textsc{Shark} is unique for reproducing the data) as well as the study of the $j_{\rm H_2}-M_{\rm H_2}$ relation at higher redshifts, for which we now provide a local baseline.

\section{Conclusions}
\label{secconclusion}
Exploiting observations from the state-of-the-art PHANGS-ALMA survey, we characterised for the first time the scaling relation between the specific angular momentum of molecular gas ($j_{\rm H_2}$) and molecular gas mass ($M_{\rm H_2}$) for disc galaxies in the local universe. Our analysis reveals a clear power-law correlation in the $j_{\rm H_2}-M_{\rm H_2}$ plane (Fig.~\ref{Fig_rela}), which we fit with a power law. The best-fitting relation has a slope of $\alpha = 0.53 \pm 0.04$ and an intercept of $\beta = 2.62 \pm 0.02$, aligning closely in slope with the well-studied stellar $j$-$M$ relation and contrasting with the steeper slope observed for neutral atomic gas (H\,{\sc i}).


We compared our findings with the predictions from an analytical model based purely on disc stability, and find that it does not recover the $j_{\rm H_2}-M_{\rm H_2}$ completely. On the other hand, compared with predictions from the \textsc{Shark} semi-analytical model, we find a better agreement (Fig.~\ref{Fig_shark}), which supports the model’s portrayal of molecular gas dynamics in galaxy evolution. 

Overall, our work provides a valuable tool for upcoming studies of gas dynamics at earlier cosmic times and places new constraints to be reproduced by galaxy formation models and simulations.

\begin{acknowledgements}
We want to thank Gabriele Pezzulli, Luca Cortese, Alessandro Romeo, Michael Fall, Francesca Rizzo, and Danail Obreschkow for valuable discussions, and Paul van der Werf for reading an earlier version of this work. We also thank the anonymous referee for a constructive report that helped improve the paper. PEMP acknowledges the support from the Dutch Research Council (NWO) through the Veni grant VI.Veni.222.364. We thank the PHANGS collaboration for making their data available. This paper makes use of the following ALMA data: \texttt{ADS/JAO.ALMA\#2012.1.00650.S}, \texttt{ADS/JAO.ALMA\#2015.1.00925.S}, \texttt{ADS/JAO.ALMA\#2015.1.00956.S}, \texttt{ADS/JAO.ALMA\#2017.1.00886.L}, \texttt{ADS/JAO.ALMA\#2018.1.01651.S}. ALMA is a partnership of ESO (representing its member states), NSF (USA) and NINS (Japan), together with NRC (Canada), MOST and ASIAA (Taiwan), and KASI (Republic of Korea), in cooperation with the Republic of Chile. The Joint ALMA Observatory is operated by ESO, AUI/NRAO and NAOJ. The National Radio Astronomy Observatory is a facility of the National Science Foundation operated under cooperative agreement by Associated Universities, Inc. We have used SIMBAD, NED, and ADS services extensively, as well as the Python packages NumPy \citep{oliphant_python_2007}, Matplotlib \citep{hunter_matplotlib_2007}, SciPy \citep{virtanen_scipy_2020}, and Astropy \citep{astropy_collaboration_astropy_2018}, for which we are thankful.
\end{acknowledgements}

%
\bibliographystyle{aa} 
\bibliography{references_copy_10mar} 
%
\begin{appendix}
 \onecolumn
 \section{Molecular angular momentum catalogue}
  \label{sec: appjvalues}
 In Table~\ref{tablejvals} we show galaxy properties and our calculated values for $j_{\rm H_2}$, $M_{\rm H_2}$, and $\mathcal{R}$ for our sample of 51 converged galaxies. Additionally, we provide the results for our non-converged sample in Table~\ref{tablenonconvjvals}, but we note  that these results are less reliable. In both tables, Hubble types, distances, and stellar masses are  from \cite{leroy_surveyphangs-alma_2021}. 

\renewcommand{\arraystretch}{1.5}
\begin{longtable}{lcccccc}
\caption{Selected galaxy properties.}\\
\hline\hline
GAL ID & Hubble type & Distance & $\log(M_{\ast})$ & $\log(j_{\mathrm{H_2}})$ & $\log(M_{\mathrm{H_2}})$ & $\mathcal{R}$ \\
    &   & (Mpc)          & ($M_\odot$)  & (kpc km s$^{-1}$)        & ($M_\odot$)        &             \\
\hline
\endfirsthead
\caption{Continued.}\\ 
\hline\hline
GAL ID & Hubble type & Distance & $\log(M_{\ast})$ & $\log(j_{\mathrm{H_2}})$ & $\log(M_{\mathrm{H_2}})$ & $\mathcal{R}$ \\
     &  & (Mpc)          & ($M_\odot$)    & (kpc km s$^{-1}$)        & ($M_\odot$)         &             \\
\hline
\endhead
\hline
\endfoot
\hline
\endlastfoot
IC 1954   & Sb       & $12.80 \pm 2.17$ & 9.67 & $2.29^{+0.07}_{-0.07}$ & $8.65^{+0.15}_{-0.13}$ & 0.98 \\
IC 5273   & SBc      & $14.18 \pm 2.14$ & 9.73 & $2.48^{+0.10}_{-0.39}$ & $8.57^{+0.14}_{-0.36}$ & 0.88 \\
NGC 253  & SABc     & $3.70 \pm 0.12$  & 10.64 & $2.77^{+0.06}_{-0.36}$ & $9.53^{+0.10}_{-0.29}$ & 1.00 \\
NGC 300  & Scd      & $2.09 \pm 0.09$  & 9.27  & $2.05^{+0.05}_{-0.04}$ & $7.67^{+0.06}_{-0.05}$ & 1.00 \\
NGC 628  & Sc       & $9.84 \pm 0.63$  & 10.34 & $2.84^{+0.07}_{-0.05}$ & $9.41^{+0.08}_{-0.07}$ & 0.79 \\
NGC 1087  & Sc       & $15.85 \pm 2.22$ & 9.94  & $2.51^{+0.06}_{-0.06}$ & $8.99^{+0.12}_{-0.11}$ & 0.92 \\
NGC 1097  & Sbb      & $13.58 \pm 2.05$ & 10.76 & $3.04^{+0.06}_{-0.13}$ & $9.62^{+0.11}_{-0.10}$ & 0.85 \\
NGC 1300  & Sbc      & $18.99 \pm 2.86$ & 10.62 & $3.09^{+0.08}_{-0.07}$ & $9.23^{+0.14}_{-0.11}$ & 0.74 \\
NGC 1317  & SABa     & $19.11 \pm 0.85$ & 10.62 & $2.12^{+0.06}_{-0.06}$ & $8.78^{+0.08}_{-0.07}$ & 0.98 \\
NGC 1365  & Sb       & $19.57 \pm 0.78$ & 11.00 & $3.17^{+0.04}_{-0.23}$ & $10.14^{+0.05}_{-0.18}$ & 1.00 \\
NGC 1385  & Sc       & $17.22 \pm 2.60$ & 9.98  & $2.47^{+0.06}_{-0.06}$ & $9.11^{+0.13}_{-0.11}$ & 1.00 \\
NGC 1511  & Sab      & $15.28 \pm 2.26$ & 9.92  & $2.60^{+0.06}_{-0.06}$ & $9.07^{+0.13}_{-0.12}$ & 0.77 \\
NGC 1546  & S0-a     & $17.69 \pm 2.02$ & 10.37 & $2.49^{+0.05}_{-0.05}$ & $9.23^{+0.10}_{-0.09}$ & 0.95 \\
NGC 1559  & SBc      & $19.44 \pm 0.45$ & 10.37 & $2.84^{+0.01}_{-0.01}$ & $9.54^{+0.02}_{-0.02}$ & 0.72 \\
NGC 1566  & SABb     & $17.69 \pm 2.02$ & 10.79 & $3.00^{+0.05}_{-0.04}$ & $9.69^{+0.10}_{-0.08}$ & 0.93 \\
NGC 1792  & Sbc      & $16.20 \pm 2.44$ & 10.62 & $2.83^{+0.06}_{-0.06}$ & $9.76^{+0.14}_{-0.12}$ & 0.89 \\
NGC 2090  & Sbc      & $11.75 \pm 0.84$ & 10.04 & $2.42^{+0.03}_{-0.03}$ & $8.47^{+0.09}_{-0.08}$ & 0.97 \\
NGC 2283  & Sc       & $13.68 \pm 2.06$ & 9.89  & $2.44^{+0.07}_{-0.07}$ & $8.55^{+0.14}_{-0.12}$ & 0.71 \\
NGC 2835  & Sc       & $12.22 \pm 0.93$ & 10.0  & $2.42^{+0.06}_{-0.06}$ & $8.42^{0.12}_{-0.11}$ & 0.79 \\
NGC 2903  & Sbc      & $10.00 \pm 1.99$ & 10.64 & $3.04^{+0.08}_{-0.08}$ & $9.53^{+0.17}_{-0.15}$ & 1.00 \\
NGC 2997  & SABc     & $14.06 \pm 2.80$ & 10.73 & $3.16^{+0.09}_{-0.10}$ & $9.80^{+0.19}_{-0.17}$ & 0.72 \\
NGC 3059  & SBbc     & $20.23 \pm 4.04$ & 10.38 & $2.65^{+0.08}_{-0.08}$ & $9.36^{+0.18}_{-0.15}$ & 0.81 \\
NGC 3137  & SABc     & $16.37 \pm 2.34$ & 9.88  & $2.48^{+0.07}_{-0.06}$ & $8.45^{+0.13}_{-0.11}$ & 0.88 \\
NGC 3351  & Sb       & $9.96 \pm 0.33$  & 10.37 & $2.88^{+0.04}_{-0.04}$ & $9.03^{+0.06}_{-0.04}$ & 1.00 \\
NGC 3507  & Sbb      & $23.55 \pm 3.99$ & 10.40 & $2.93^{+0.08}_{-0.06}$ & $9.17^{+0.15}_{-0.13}$ & 0.73 \\
NGC 3511  & SABc     & $13.94 \pm 2.10$ & 10.03 & $2.70^{+0.06}_{-0.06}$ & $8.99^{+0.13}_{-0.11}$ & 0.77 \\
NGC 3521  & SABb     & $13.24 \pm 1.96$ & 11.003 & $3.12^{+0.06}_{-0.06}$ & $9.75^{+0.13}_{-0.11}$ & 1.00 \\
NGC 3596  & SABc     & $11.30 \pm 1.03$ & 9.66  & $2.32^{+0.07}_{-0.06}$ & $8.61^{+0.12}_{-0.10}$ & 0.91 \\
NGC 3621  & SBcd     & $7.06 \pm 0.28$   & 10.06 & $2.68^{+0.06}_{-0.05}$ & $9.06^{+0.06}_{-0.06}$ & 1.00 \\
NGC 3626  & S0-a     & $20.05 \pm 2.34$  & 10.46 & $2.28^{+0.05}_{-0.05}$ & $8.35^{+0.14}_{-0.11}$ & 0.98 \\
NGC 3627  & S0-a     & $11.32 \pm 0.48$  & 10.84 & $3.02^{+0.05}_{-0.06}$ & $9.78^{+0.13}_{-0.11}$ & 0.98 \\
NGC 4254  & Sc       & $13.10 \pm 2.01$  & 10.42 & $2.90^{+0.07}_{-0.06}$ & $9.80^{+0.13}_{-0.12}$ & 0.97 \\
NGC 4293  & S0-a     & $15.76 \pm 2.38$  & 10.52 & $2.21^{+0.08}_{-0.08}$ & $8.63^{+0.15}_{-0.13}$ & 0.79 \\
NGC 4298  & Sc       & $14.92 \pm 1.36$  & 10.04 & $2.55^{+0.05}_{-0.04}$ & $9.10^{+0.09}_{-0.08}$ & 0.96 \\
NGC 4303  & Sbc      & $16.99 \pm 3.02$  & 10.51 & $2.91^{+0.07}_{-0.08}$ & $9.78^{+0.16}_{-0.15}$ & 0.81 \\
NGC 4457  & S0-a     & $15.10 \pm 2.00$  & 10.42 & $2.41^{+0.08}_{-0.07}$ & $8.82^{+0.16}_{-0.13}$ & 0.72 \\
NGC 4496A & Scd      & $14.86 \pm 1.06$  & 9.55  & $2.28^{+0.08}_{-0.09}$ & $8.38^{+0.11}_{-0.11}$ & 0.71 \\
NGC 4535  & Sc       & $15.77 \pm 0.37$  & 10.54 & $3.06^{+0.04}_{-0.04}$ & $9.44^{+0.06}_{-0.05}$ & 0.98 \\
NGC 4536  & SABb     & $16.25 \pm 1.12$  & 10.40 & $2.82^{+0.08}_{-0.07}$ & $9.33^{+0.10}_{-0.08}$ & 0.99 \\
NGC 4540  & SABc     & $15.76 \pm 2.38$  & 9.79  & $2.23^{+0.08}_{-0.07}$ & $8.50^{+0.15}_{-0.12}$ & 0.89 \\
NGC 4689  & Sc       & $15.00 \pm 2.26$  & 10.24 & $2.50^{+0.07}_{-0.06}$ & $9.04^{+0.14}_{-0.11}$ & 0.92 \\
NGC 4781  & Scd      & $11.31 \pm 1.18$  & 9.54  & $2.34^{+0.06}_{-0.06}$ & $8.69^{+0.12}_{-0.10}$ & 0.99 \\
NGC 4941  & SABa     & $15.00 \pm 5.00$  & 10.18 & $2.55^{+0.19}_{-0.12}$ & $8.37^{+0.41}_{-0.12}$ & 0.99 \\
NGC 4951  & SABc     & $15.00 \pm 4.19$  & 9.79  & $2.36^{+0.12}_{-0.11}$ & $8.40^{+0.25}_{-0.20}$ & 0.94 \\
NGC 5042  & SABc     & $16.78 \pm 2.53$  & 9.90  & $2.53^{+0.08}_{-0.07}$ & $8.52^{+0.15}_{-0.12}$ & 0.79 \\
NGC 5134  & SABb     & $19.92 \pm 2.69$  & 10.41 & $2.58^{+0.08}_{-0.07}$ & $8.76^{+0.13}_{-0.11}$ & 0.80 \\
NGC 5248  & SABb     & $14.87 \pm 1.32$  & 10.41 & $2.83^{+0.05}_{-0.05}$ & $9.50^{+0.08}_{-0.08}$ & 0.72 \\
NGC 5530  & SABb     & $12.27 \pm 1.85$  & 10.08 & $2.52^{+0.07}_{-0.06}$ & $8.77^{+0.14}_{-0.12}$ & 0.89 \\
NGC 5643  & Sc       & $12.68 \pm 0.54$  & 10.34 & $2.80^{+0.03}_{-0.03}$ & $9.41^{+0.05}_{-0.05}$ & 0.88 \\
NGC 6300  & Sbb      & $11.58 \pm 1.75$  & 10.47 & $2.74^{+0.06}_{-0.06}$ & $9.25^{+0.13}_{-0.12}$ & 0.88 \\
NGC 7793  & Scd      & $3.62 \pm 0.15$   & 9.36  & $2.22^{+0.07}_{-0.09}$ & $8.18^{+0.11}_{-0.12}$ & 1.00 \\
\label{tablejvals}
\end{longtable}

\renewcommand{\arraystretch}{1.5}
\begin{longtable}{lcccccc}
\caption{Selected galaxy properties non-converged sample.}\\
\hline\hline
GAL ID & Hubble Type & Distance & $\log(M_{\ast})$ & $\log(j_{\mathrm{H_2}})$ & $\log(M_{\mathrm{H_2}})$ & $\mathcal{R}$ \\
   &    & (Mpc)          & ($M_\odot$)   & (kpc km s$^{-1}$)        & ($M_\odot$)         &             \\
\hline
\endfirsthead
\caption{Continued.}\\ 
\hline\hline
GAL ID & Hubble Type & Distance & $\log(M_{\ast})$ & $\log(j_{\mathrm{H_2}})$ & $\log(M_{\mathrm{H_2}})$ & $\mathcal{R}$ \\
    &   & (Mpc)          & ($M_\odot$)    & (kpc km s$^{-1}$)        & ($M_\odot$)         &             \\
\hline
\endhead
\hline
\endfoot
\hline
\endlastfoot
NGC 685 & Sc & $19.94 \pm 3.01$ & 10.07 & $2.77^{+0.08}_{-0.09}$ & $8.74^{+0.11}_{-0.14}$ & 0.68 \\
NGC 1433 & SBa   & $18.63 \pm 1.84$ & 10.87 & $3.13^{+0.08}_{-0.10}$ & $9.30^{+0.08}_{-0.10}$ & 0.25 \\
NGC 1809 & Sc  & $19.95 \pm 5.63$ & 9.77 & $2.06^{+0.11}_{-0.13}$ & $8.13^{+0.20}_{-0.26}$ & 0.67 \\
NGC 4321 & SABb  & $15.21 \pm 0.50$ & 10.75  & $3.21^{+0.05}_{-0.06}$ & $9.96^{+0.06}_{-0.07}$ & 0.53 \\
NGC 4548 & Sb & $16.22 \pm 0.38$  & 10.70 & $3.30^{+0.07}_{-0.11}$ & $9.40^{+0.10}_{-0.18}$ & 0.38 \\
NGC 4571 & Sc  & $14.90 \pm 1.07$  & 10.10 & $2.89^{+0.09}_{-0.31}$ & $9.17^{+0.11}_{-0.41}$ & 0.59 \\
NGC 6744 & Sbc & $9.39 \pm 0.42$  & 10.72  & $3.26^{+0.03}_{-0.03}$ & $9.62^{+0.04}_{-0.04}$ & 0.58 \\
NGC 7456 & Sc  & $15.70 \pm 2.33$ & 9.65 & $2.50^{+0.08}_{-0.09}$ & $7.98^{+0.09}_{-0.13}$ & 0.44 \\
NGC 7496 & Sb & $18.72 \pm 2.82$ & 10.00 & $2.50^{+0.06}_{-0.07}$ & $9.11^{+0.13}_{-0.14}$ & 0.67 \\
\label{tablenonconvjvals}
\end{longtable}

 \twocolumn

\section{Including H\,{\sc i} rotation curves}
\label{apprc}
\begin{figure}[h!]
   \centering
\includegraphics[width=\hsize]{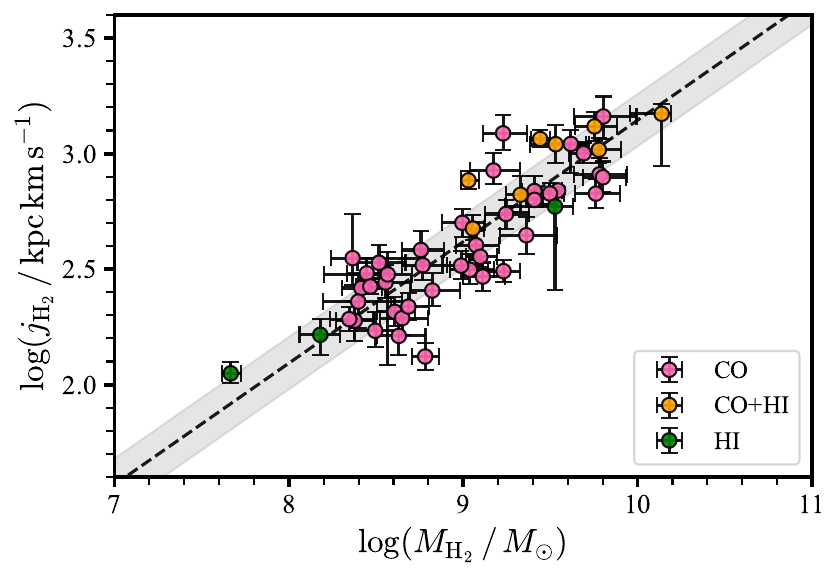}
      \caption{Converged sample of 51 galaxies in the $j_{\rm H_2}-M_{\rm H_2}$ plane. We indicate whether the CO rotation curve (pink), the CO curve supplemented with the H\,{\sc i} rotation curve (orange), or only the H\,{\sc i} rotation curve (green) was used to determine $j_{\rm H_2}$ for each galaxy.}
         \label{Fig_HI}
   \end{figure}

To extend the radial coverage of our sample’s kinematics, we supplemented our CO rotation curves with literature H\,{\sc i} rotation curves when available (see Sect.\ref{sec:rotationcurves}). In Fig.\ref{Fig_HI} we highlight in the $j_{\rm H_2}-M_{\rm H_2}$ plane \emph{(i)} the eight galaxies whose CO rotation curves are complemented by H\,{\sc i} velocities (orange), \emph{(ii)} the three galaxies for which only H\,{\sc i} data were used (green), and \emph{(iii)} all remaining galaxies that rely solely on CO rotation curves (pink).

We tested how including H\,{\sc i} rotation curves could affect the $j_{\rm H_2}-M_{\rm H_2}$ relation by examining the influence of these galaxies on the best-fit parameters (Table~\ref{tabRCslopes}). Any resulting changes in the slope lie within the uncertainties of our fiducial relation, confirming the robustness of our approach and indicating that the inclusion of H\,{\sc i} rotation curves does not bias our results presented in Sect.~\ref{sec:jmrela}.

\begin{table}[h!]
\caption{Impact of rotation curves on the best-fit parameters to the $j_{\rm H_2}-M_{\rm H_2}$ relation.}             
\label{tabRCslopes}      
\centering  
\resizebox{\columnwidth}{!}{
\begin{tabular}{c c c c c}       
\hline\hline                 
 Rotation curves & N galaxies & $\alpha$ & $\beta$ & $\sigma_{\perp}$\\    
\hline                       
   Fiducial & 51 & $0.53^{+0.04}_{-0.04}$ & $2.62^{+0.02}_{-0.02}$   & $0.11^{+0.02}_{-0.02}$\\      
   CO \& CO+H\,{\sc i} & 48 & $0.56^{+0.05}_{-0.05}$ & $2.61^{+0.02}_{-0.02}$   & $0.12^{+0.02}_{-0.02}$\\
   CO \& H\,{\sc i}& 43 & $0.49^{+0.05}_{-0.05}$ & $2.60^{+0.02}_{-0.02}$   & $0.11^{+0.02}_{-0.02}$ \\
   CO & 40 & $0.52^{+0.06}_{-0.05}$ & $2.60^{+0.02}_{-0.02}$   & $0.11^{+0.02}_{-0.02}$ \\ 
\hline                                   
\end{tabular} }
\label{tabRCslopes}
\end{table}
\section{Convergence criteria}
\label{sec: appconvergence}
We obtain the molecular $j-M$ relation by imposing a convergence limit $\mathcal{R} > 0.7$, as detailed in Sec.~\ref{secconv}. This appendix explores how robust our findings are against our $\mathcal{R}$ threshold choice.

\begin{figure}[h!]
   \centering
\includegraphics[width=\hsize]{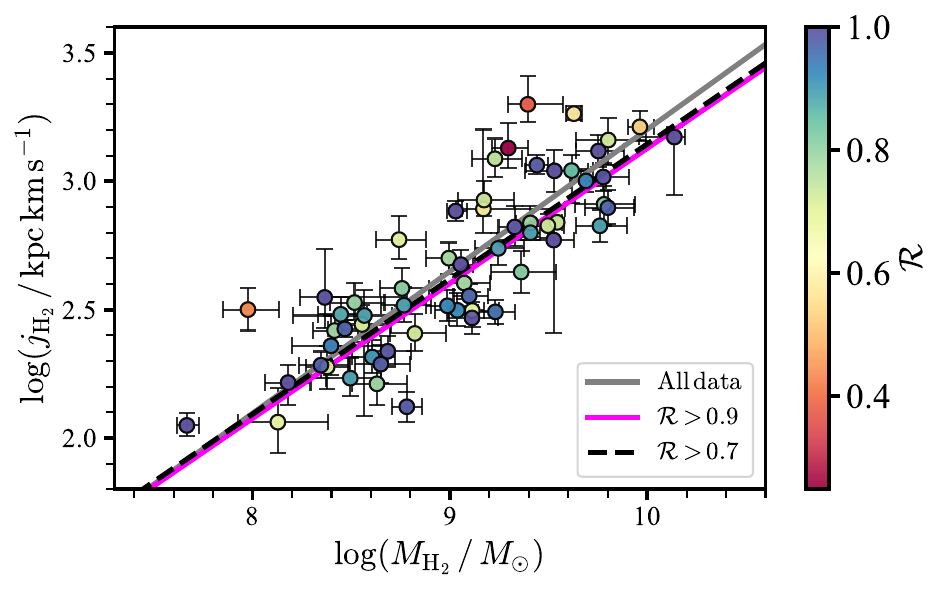}
      \caption{Total sample of 60 galaxies in the $j_{\rm H_2}-M_{\rm H_2}$ plane, with their convergence factors, as given in Table \ref{tablejvals}, indicated by colour. The best-fitting relations to samples with different $\mathcal{R}$ thresholds are shown with grey (all data), magenta ($\mathcal{R}$ > 0.9), and black ($\mathcal{R}$ > 0.7 ) curves.
              }
         \label{Fig_conv}
   \end{figure}

Figure \ref{Fig_conv} shows the total sample of 60 galaxies in the $j_{\rm H_2}-M_{\rm H_2}$ plane with $\mathcal{R}$ indicated by the colour of the markers, the values of the non-converged galaxies are provided in Table~\ref{tablenonconvjvals}. To test the impact of our $\mathcal{R}$ threshold, we repeat our fitting procedure but include the galaxies satisfying $\mathcal{R}>0.2$ (in which our full sample is comprised) and the more restrictive $\mathcal{R}>0.9$. The figure shows the resulting best-fit relations, and the fitting parameters are provided in Table~\ref{tabslopes}. Typically, the galaxies with the lowest $\mathcal{R}$ lie above the fiducial relation, which may be counterintuitive, but it is likely the result of the functional forms overestimating the velocity at which the rotation curve flattens. 

As can be seen from Fig.~\ref{Fig_rela} and Table~\ref{tabslopes}, all the fitting parameters remain consistent within their uncertainties regardless of the specific convergence criterion employed. Even though the sample size decreases as we impose stricter convergence limits, we still recover slopes and intercepts that are statistically consistent. This all demonstrates the robustness of our approach and results.  

\begin{table}[h!]
\caption{Best-fit parameters to the $j_{\rm H_2}-M_{\rm H_2}$ relation, varying the minimum required $\mathcal{R}$. }             
\label{tableslopes}      
\centering                         
\begin{tabular}{c c c c c}       
\hline\hline                 
 $\mathcal{R_{\rm min}}$ & N galaxies & $\alpha$ & $\beta$ & $\sigma_{\perp}$\\    
\hline                       
   $0.2$ & 60 & $0.55^{+0.05}_{-0.05}$ & $2.65^{+0.02}_{-0.02}$   & $0.14^{+0.02}_{-0.02}$\\      
   $0.7$ & 51 & $0.53^{+0.04}_{-0.04}$ & $2.62^{+0.02}_{-0.02}$   & $0.11^{+0.02}_{-0.02}$\\
   $0.8$& 37 & $0.52^{+0.05}_{-0.05}$ & $2.60^{+0.03}_{-0.03}$   & $0.12^{+0.02}_{-0.02}$ \\
   $0.9$ & 26 & $0.54^{+0.06}_{-0.06}$ & $2.60^{+0.03}_{-0.04}$   & $0.14^{+0.03}_{-0.02}$ \\ 
\hline                                   
\end{tabular}
\label{tabslopes}
\end{table}

\section{The $j_{\rm H_2}-M_{\rm \ast}$ relation}
\label{appstellar}
\begin{figure}[h!]
   \centering
   \includegraphics[width=\hsize]{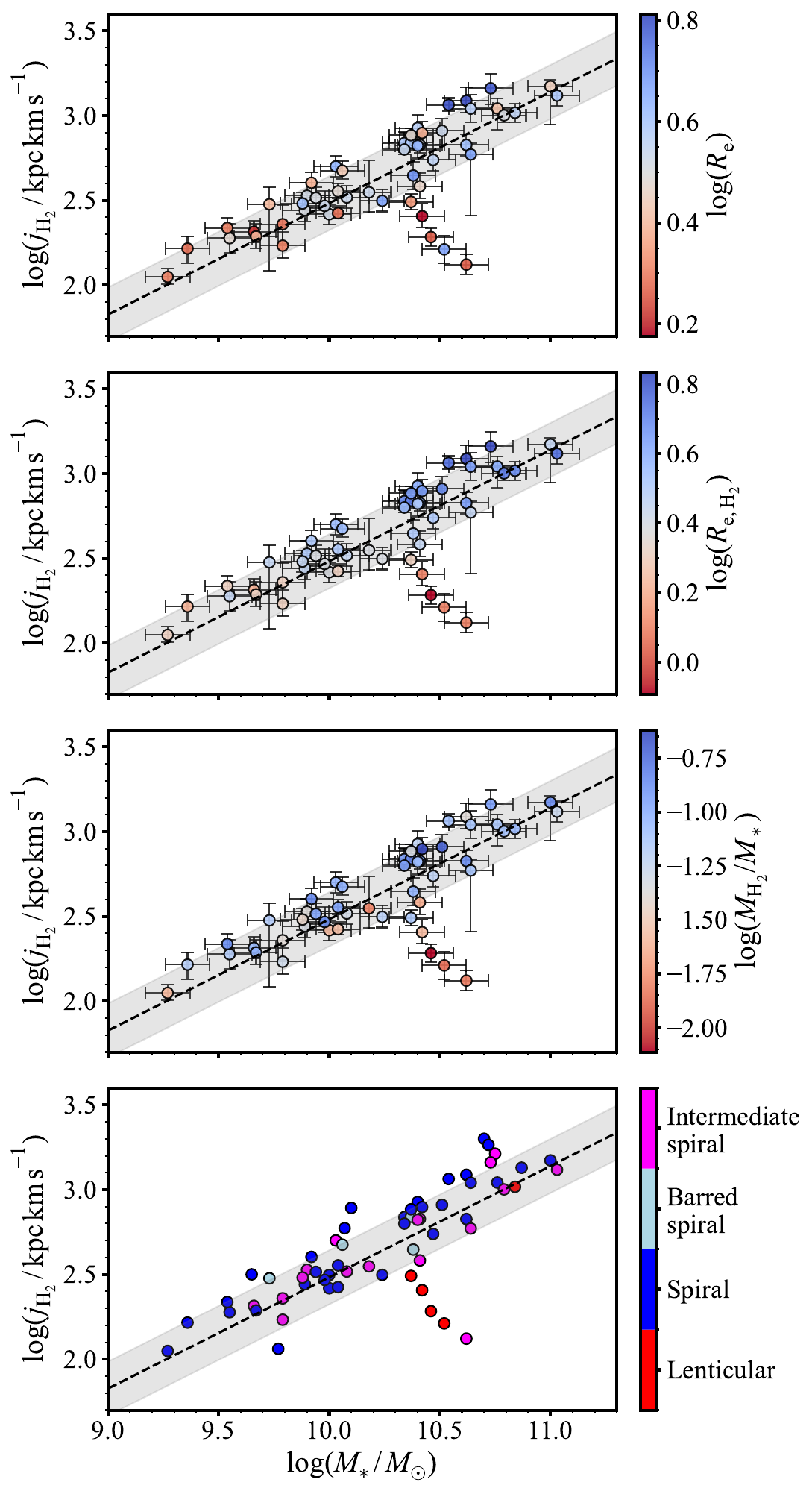}
   \caption{$j_{\rm H_2}-M_{\ast}$ relation for our converged sample, the dashed line and grey band indicate the best-fitting relation and its orthogonal intrinsic scatter, respectively. The panels highlight the dependence of the relation on the effective radius (top),  $R_{\rm e, H_2}$, the radius within which half of $M_{\rm H_2}$ is contained (upper-middle), molecular gas fraction (lower-middle), and morphology (bottom).}
   \label{Figappendixb}
\end{figure}

Our analysis also allows us to explore the $j_{\rm H_2}$ content at fixed $M_\ast$, the dominant baryonic mass component for all our galaxies. The stellar masses are provided in Table~\ref{tablejvals}, where we adopt an uncertainty of 0.1 dex \citep{leroy_surveyphangs-alma_2021}. 
As shown in Fig.~\ref{Figappendixb}, except for a few outliers, most of our converged sample follows a clear trend in the $j_{\rm H_2}-M_{\rm \ast}$ plane, which we parametrise with the relation
\begin{align}
\log\left( \frac{j_{\rm H_2}}{\text{kpc km s}^{-1}} \right) &= \alpha \left[ \log\left( \frac{M_{\ast}}{M_{\odot}} \right) - 10 \right] + \beta~.
\label{eq: powerlawfitstar}
\end{align}
This expression differs from Eq.~\ref{eq: powerlawfit} only in the mass shift (which affects the normalisation $\beta$). The best-fitting parameters are $\alpha = 0.66 \pm 0.08$, $\beta = 2.48 \pm 0.03$, with an orthogonal intrinsic scatter $\sigma_{\perp}=0.16 \pm 0.02$. We have checked that our fit is not significantly affected by the presence of the four outliers (see below): excluding these four galaxies does not affect the slope of either the $j_{\rm H_2}$–$M_{\rm H_2}$ or the $j_{\rm H_2}$–$M_\ast$ relations, indicating that our results are robust.

As for the $j_{\rm H_2}-M_{\rm H_2}$ relation, we look for second dependences in the $j_{\rm H_2}-M_{\ast}$ plane. In contrast with the $j_{\rm H_2}-M_{\rm H_2}$ relation (where we find an anti-correlation), we find a strong positive correlation with $M_{\rm H_2}/M_\ast$ ($p$-value = 0.00005).
Furthermore, we find a significant ($p$-value = 0.0004) correlation with $\log(R_{\rm e})$, indicating that galaxies with more extended light distribution tend to have higher $j_{\rm H_2}$, as also observed for $j_\ast$ \citep{mancera_pina_baryonic_2021}. We show the dependences on $\log(R_{\rm e})$ and $M_{\rm H_2}/M_\ast$, in the top and lower-middle panel of Fig.~\ref{Figappendixb}.

 Additionally, we compute $R_{\rm e, H_2}$, the radius within which half of the molecular gas is contained. This quantity, analogous to the stellar effective radius $R_{\rm e}$, is derived from integrating the cumulative $\Sigma_{\rm H_2}$ profiles. Similar to $R_{\rm e}$, we find a robust positive correlation between $j_{\rm H_2}$ and $\log(R_{\rm e, H_2})$ with a $p$-value of $2 \times 10^{-9}$.
This molecular size dependence is seen across our mass regime, but it is particularly evident at $\log(M_{\ast}/M_\odot)\approx10.5$, where we observe a tail of outliers exhibiting significantly lower $R_{\rm e, H_2}$ and $j_{\rm H_2}$, indicating that their molecular gas is significantly less extended that for the other galaxies\footnote{The outliers are extreme but do not solely drive the correlation; significant $p-$values for $M_{\rm H_2}/M_\ast$ ($p$-value = 0.003), $\log(R_{\rm e})$ ($p$-value = 0.004), and $\log(R_{\rm e, H_2})$ ($p$-value = 0.00005), remain without the tail.}.
The tail is formed by the galaxies NGC~1317, NGC~3626, NGC~4293, and NGC~4457. 

Upon inspection, we find that in addition to their very low  $R_{\rm e, H_2}$, these galaxies show significantly lower molecular gas fractions, as illustrated in the lower-middle panel of Fig.~\ref{Figappendixb}. We also notice that none of these galaxies is a normal spiral; instead, they are classified as lenticulars (S0) or intermediate spirals (SABa) based on their morphology (bottom panel of  Fig.~\ref{Figappendixb}, Hubble types are provided in Table~\ref{tablejvals}). This was not necessarily expected, as the discs of lenticular galaxies (and we have checked this is the case for our sample) are found to lie on the same $j_\ast-M_\ast$ relation for spirals, as shown by \cite{rizzo_s0_2018, mancera_pina_baryonic_2021}, see also discussion in \cite{romanowsky_angular_2012}.

The combination of normal $j_\ast$ but low $j_{\rm H_2}$, $R_{\rm e, H_2}$, and $M_{\rm H_2}/M_\ast$, seen primarily for lenticular or intermediate spirals, is likely related to their formation mechanism, which remains a topic of active study \citep{deeley_sami_2020}. These mechanisms include passive evolution, where cold gas is gradually consumed by star formation \citep{van_den_bergh_lenticular_2009, laurikainen_photometric_2010, bellstedt_sluggs_2017}, or more violent processes such as mergers and tidal interactions that rapidly strip or deplete the gas \citep{bekki_transformation_2011, falcon-barroso_angular_2015, querejeta_formation_2015}. 
Whatever the exact formation channel  at play, our results suggest that lenticular galaxies, for example  spirals, self-regulate to lie on the $j_\ast-M_\ast$ and $j_{\rm H_2}-M_{\rm H_2}$ relations (see also \citealt{mancera_pina_tight_2021}) but have gone through a process that altered mostly their gas disc, preferentially removing gas from the outer, high-$j_{\rm H_2}$ regions. As a result, $j_{\rm H_2}$ and $M_{\rm H_2}$ are significantly reduced, while the dominant stellar components remain largely unaffected. These observations suggest a variety of formation paths for lenticulars,  or at least diverse evolutionary paths for their molecular gas reservoirs, which may explain the large observed range of $j_{\rm H_2}$ at fixed $M_{\ast}$ for this galaxy type. This highlights the importance of angular momentum studies in understanding the diversity in galaxy morphology observed in the Universe. However, considering that our sample is predominantly composed of galaxies with minimal bulge components, a more representative sample, including a greater number of lenticular galaxies, would be necessary to draw  clearer conclusions.

\end{appendix}

\end{document}